\documentclass[reprint,amsmath,amssymb,aps,prb]{revtex4-2}
\usepackage{graphicx}
\usepackage{amsmath}
\usepackage{bm}
\usepackage[mathlines]{lineno}
\usepackage{xcolor}
\definecolor{blue}{RGB}{0,0,255}
\usepackage[colorlinks=true,linkcolor=blue,citecolor=blue,urlcolor=blue,]{hyperref}
\UseRawInputEncoding
\usepackage{booktabs} 
\usepackage{gensymb} 
\usepackage{amsmath, nccmath, graphicx}
\usepackage{afterpage}
\usepackage{bm}

\usepackage{times}
\usepackage{graphicx}
\usepackage{float}
\usepackage{latexsym,amsmath,amssymb,bm,euscript}
\usepackage{color}
\usepackage{subfigure}
\usepackage{epstopdf}
\usepackage[colorlinks=true,linkcolor=blue,citecolor=blue]{hyperref}
\usepackage{soul}
\usepackage[normalem]{ulem}
\usepackage{mathrsfs}
\usepackage{amsmath}
\usepackage{lettrine}
\usepackage{bbding}
\usepackage{xspace}
\usepackage{textcomp}
\usepackage{textcase}
\usepackage{setspace}
\usepackage{multirow}
\usepackage[version=4]{mhchem}

\begin{document}
	\preprint{APS/123-QED}
	\title{{\color{black}Field-induced multipolar character in the dipolar ground state of the honeycomb rare-earth chalcohalide NdOF}}

    \author{Tiantian\,Liu\textsuperscript{\textcolor{blue}{$^{1,2}$}}}
    \author{Yanzhen\,Cai\textsuperscript{\textcolor{blue}{$^{1,2}$}}}
    \author{Mingtai\,Xie\textsuperscript{\textcolor{blue}{$^{1,2}$}}}
    \author{Helin\,Mei\textsuperscript{\textcolor{blue}{$^{1,2}$}}}
    \author{Anmin\,Zhang\textsuperscript{\textcolor{blue}{$^{1}$}}}
    \author{Feng\,Jin\textsuperscript{\textcolor{blue}{$^{2}$}}}
    \author{Jianting\,Ji\textsuperscript{\textcolor{blue}{$^{2}$}}}
    \author{Zheng\,Zhang\textsuperscript{\textcolor{blue}{$^{2}$}}}
    \email{zhangzheng@iphy.ac.cn}
    \author{Qingming\,Zhang\textsuperscript{\textcolor{blue}{$^{2,1}$}}}
    \email{qmzhang@iphy.ac.cn}

\affiliation{$^{1}$School of Physical Science and Technology, Lanzhou University, Lanzhou 730000, China}
\affiliation{$^{2}$Beijing National Laboratory for Condensed Matter Physics, Institute of Physics, Chinese Academy of Sciences, Beijing 100190, China}
	
\begin{abstract}
    {\color{black}Field-tunable reconstruction of crystalline electric field (CEF) doublets offers a promising avenue for inducing multipolar character, while its observation in real materials has been little explored so far.
    	Here we establish the honeycomb rare-earth chalcohalide NdOF as such a platform. 
    	Raman spectroscopy identifies four CEF excitations at 1.7, 15.6, 19.2, and 80.9~meV, and a Zeeman--CEF analysis reproduces their nonlinear field splitting into seven branches. 
    	{\color{black}Magnetization and susceptibility over 0.1--9~T are well described by a CEF model for the total angular momentum $J = 9/2$ manifold, confirming the robustness of the extracted CEF scheme.}
    	These results demonstrate a field-driven continuous evolution of the ground-state doublet from dipolar to dipolar-multipolar character, with pressure providing a complementary tuning knob, establishing NdOF as a model system for exploring the controlled induction of multipolar components in rare-earth magnets.}
\end{abstract}

\maketitle
\textit{Introduction} --- {\color{black}For rare-earth ions with half-integer total angular momentum $J$, the combined effect of strong spin--orbit coupling (SOC) and the crystalline electric field (CEF) produces a local ground-state Kramers doublet at each site, which dictates the low-temperature magnetic properties of the system~\cite{PhysRevB.95.085132,PhysRevB.98.134437,10.21468/SciPostPhysCore.3.1.004}. 
Any Kramers doublet can be represented by a time-reversal-odd pseudospin operator $\tau^\mu$ ($\mu=x,y,z$), but the way in which these pseudospins transform under space-group symmetries depends on the specific doublet wave functions~\cite{abragam2012electron, PhysRevLett.112.167203,PhysRevB.94.035107,PhysRevB.94.201114}.
The most common case is that all components of $\tau^\mu$ transform as magnetic dipoles, giving rise to conventional Ising or XY anisotropy. 
More exotic situations arise when different components of $\tau^\mu$ transform differently --- for example, with $\tau^z$ and $\tau^x$ behaving as dipoles while $\tau^y$ transforms as a component of a magnetic octupole tensor~\cite{PhysRevLett.112.167203}. 
This dipolar-octupolar duality within a single doublet has been shown to generate a remarkably rich theoretical phase diagram, ranging from multipolar quantum spin liquids~\cite{PhysRevB.107.L020408,shen2019intertwined,PhysRevB.98.045119,PhysRevB.108.094418,7k9l-j1kh,PhysRevB.111.L180405} to octupolar quantum spin ice~\cite{PhysRevLett.112.167203,rau2019frustrated,RevModPhys.82.53,PhysRevLett.129.097202,PhysRevB.110.174441,PhysRevResearch.2.023253} in pyrochlore materials. 
Understanding and, more importantly, controlling the dipolar versus multipolar nature of such doublets is therefore a central issue in the study of rare-earth quantum magnets.
}

\begin{figure*}[t]
	\centering
	\includegraphics[width=\linewidth]{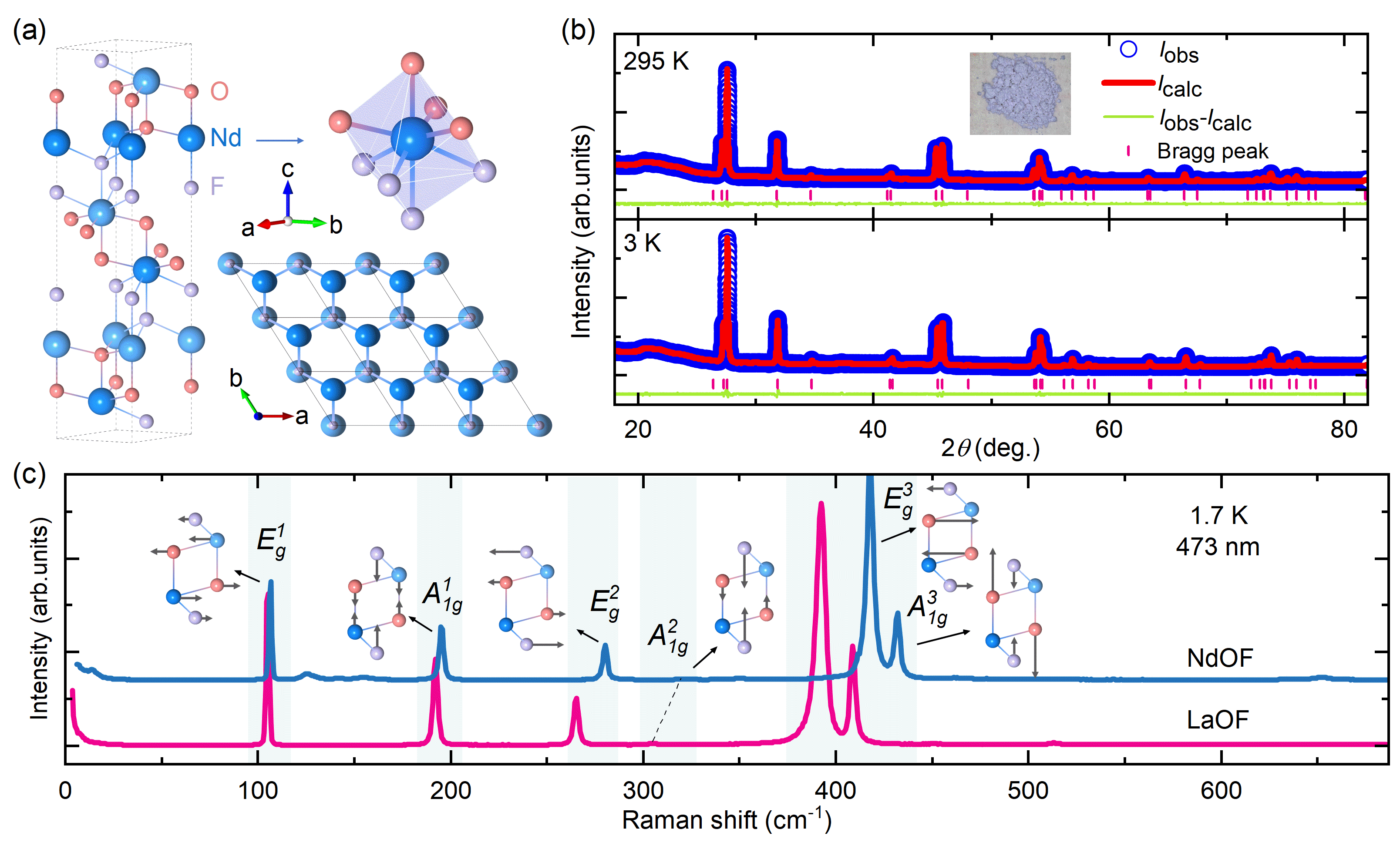}
	\caption{
		\textbf{Crystal structure, phase purity, and lattice vibrations of NdOF.}
		\textbf{(a)} Crystal structure of NdOF, which adopts the SmSI-type trigonal structure ($R\overline{3}m$). The \ce{Nd^{3+}} ion (blue) is coordinated by four O (red) and four F (purple) atoms forming a bicapped trigonal 
		antiprism with local $C_{3v}$ symmetry. The bottom panel shows the two-dimensional honeycomb arrangement of \ce{Nd^{3+}} ions in the $ab$ plane.  
		\textbf{(b)} Rietveld refinement of powder X-ray diffraction data collected at 295~K and 3~K. 
		The observed (blue circles), calculated (red line), and difference (green line) profiles are shown, with vertical bars marking the Bragg-peak positions. The results confirm the single-phase nature of NdOF and the absence of structural transitions down to 3~K. 
		\textbf{(c)} Raman spectra of NdOF and nonmagnetic reference LaOF measured at 1.7~K with 473~nm excitation. 
		Six well-defined phonon modes are identified at 106.2, 194.8, 279.8, 319.5, 417.4, and 431.6~cm$^{-1}$, corresponding to the $A_{1g}^{1}$, $E_{g}^{1}$, $A_{1g}^{2}$, $E_{g}^{2}$, $A_{1g}^{3}$, and $E_{g}^{3}$ vibrational symmetries, respectively.
	}
	\label{fig:figure1}
\end{figure*}  

{\color{black}A promising route to realize and tune such unconventional Kramers doublets is offered by layered rare-earth chalcohalides $RE$O$X$ ($RE = $ Rare Earth, $X = $ F, Cl, Br)~\cite{Ji_2021,PhysRevResearch.6.043061,PhysRevResearch.4.033006,PhysRevResearch.6.033274,rfly-9g6x}, which crystallize in the SmSI-type trigonal structure with space group $R\overline{3}m$. 
In these compounds, the rare-earth ions form robust two-dimensional honeycomb networks with local $C_{3v}$ symmetry, providing a clean structural environment for investigating CEF physics. 
The absence of structural transitions down to low temperatures further ensures that the observed excitations can be unambiguously attributed to electronic origins.
Among this family, NdOF is particularly appealing: the \ce{Nd^{3+}} ion carries a $J=9/2$ ground multiplet with a small CEF gap to the first excited doublet, rendering its ground-state wave function especially susceptible to external perturbations. 
These features make NdOF an ideal platform to explore how external fields or lattice distortions can drive a crossover from a purely dipolar doublet to one with substantial multipolar character.
}

{\color{black}In this work, we combine Raman spectroscopy, powder X-ray diffraction, and magnetization measurements to establish the CEF scheme of NdOF and its field evolution. 
Raman spectra resolve four well-defined CEF excitations within the $^{4}I_{9/2}$ manifold, including a low-lying mode at 1.7~meV, that sets a small gap to the first excited doublet. 
By developing a powder-averaged CEF analysis that incorporates Zeeman splitting, we quantitatively reproduce the nonlinear field-induced splitting of the excitations into seven distinct branches, most pronounced for in-plane fields. 
{\color{black}Consistently, the temperature-dependent magnetization ($M$-$H$) and susceptibility ($M/H$-$T$) are well reproduced by the CEF model for the total angular momentum $J = 9/2$ manifold, providing an independent verification of the reliability of the extracted CEF parameters.}
Together these results demonstrate that NdOF hosts a dipolar CEF ground doublet which, owing to the small CEF gap, can be continuously reconstructed under external fields to acquire multipolar components.
}

{\color{black}Our findings establish NdOF as a typical example of a rare-earth honeycomb magnet in which the CEF ground state, while predominantly dipolar at zero field, continuously evolves to acquire multipolar components under external fields. 
The demonstrated powder-averaged CEF framework provides a quantitative basis for describing this evolution, and the small CEF gap makes NdOF exceptionally sensitive to both magnetic fields and lattice distortions. 
This tunability highlights NdOF as a model platform for investigating dipolar-multipolar admixture, providing a solid foundation for future studies on anisotropic exchange interactions and {\color{black}the controlled induction of multipolar components in rare-earth magnets.}
}

\textit{Crystal {\color{black}structure} and Raman active phonons} ---
{\color{black}NdOF adopts the SmSI-type trigonal structure ($R\overline{3}m$), where each \ce{Nd^{3+}} ion is coordinated by four O and four F atoms in a bicapped trigonal antiprism with local $C_{3v}$ symmetry [Fig.~\ref{fig:figure1}(a)]. 
This arrangement generates quasi-two-dimensional layers of \ce{Nd^{3+}} ions forming a honeycomb network. 
Rietveld refinement of powder XRD at 295 and 3~K [Fig.~\ref{fig:figure1}(b)] confirms a pure single phase in the $R\overline{3}m$ structure without any structural 
transition (see Supplementary Materials (SM)~\cite{SuppInfo} (including references ~\cite{rietveld1969profile,rodriguez1993recent})). 
}
   
{\color{black}Lattice vibrations associated with the crystal structure can be probed by Raman scattering. Fig.~\ref{fig:figure1}\textbf{(c)} shows the low-temperature ($T=1.7$~K) Raman spectra of NdOF together with nonmagnetic LaOF, an isostructural reference compound adopting the same $R\overline{3}m$ 
symmetry. 
In LaOF, the phonon features have been well documented~\cite{holsa1997characterization,holsa1993ir}, and the absence of $4f$ magnetism allows an unambiguous assignment of the vibrational modes. 
For NdOF, six sharp phonon peaks are resolved at 106.2, 194.8, 279.8, 319.5, 417.4, and 431.6~cm$^{-1}$, corresponding to the $E_g^1$, $A_{1g}^1$, $E_g^2$, $A_{1g}^2$, $E_g^3$, and $A_{1g}^3$ modes, respectively.} {\color{black}Symmetry analysis tells us that the $A_{1g}$ mode is visible only in the parallel polarization configuration (XX), while the $E_g$ mode can be observed in XX and cross polarization configurations (XY). The two modes can be clearly identified by the polarized Raman scattering spectra. To further elucidate this, we performed an analysis based on density functional theory (DFT) using VIENNA AB INITIO SIMULATION PACKAGE (VASP) and PHONONPY package  \cite{PhysRevB.54.11169,kresse1996efficiency,togo2023implementation,togo2023first,PhysRevLett.77.3865,RevModPhys.73.515}. The corresponding vibrational modes are illustrated in Fig.~\ref{fig:figure1}\textbf{(c)} (see SM~\cite{SuppInfo}).} Additional spectra measured with different laser wavelengths (see SM~\cite{SuppInfo}) confirm that these features are intrinsic phonons rather than fluorescence artifacts.

  \begin{figure*}[t]
  	\centering
  	\includegraphics[width=\linewidth]{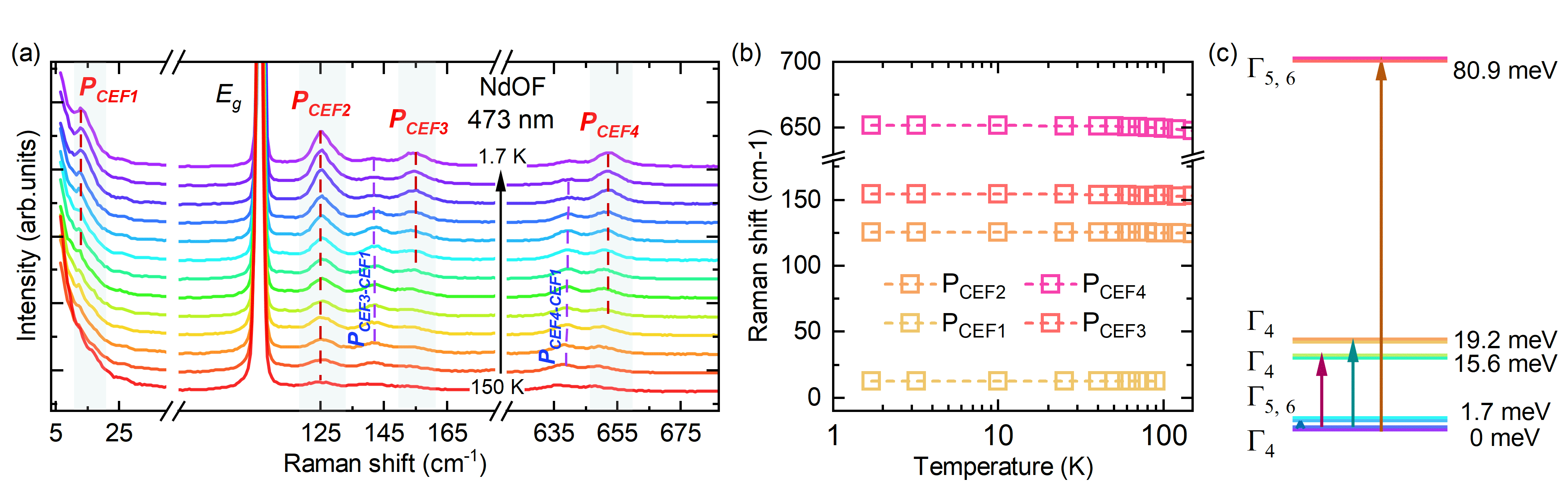}
  	\caption{ 
  	\textbf{CEF excitations of NdOF revealed by Raman scattering. }
  	\textbf{(a)} Temperature-dependent Raman spectra measured between 150~K and 1.7~K with 473~nm excitation. 
  	At low temperature four sharp intrinsic CEF peaks ($P_{\rm CEF1}$ -- $P_{\rm CEF4}$) are observed, corresponding to transitions from the ground-state doublet. 
  	Additional weaker non-intrinsic features arising from transitions between thermally populated excited doublets are visible at elevated temperatures but vanish upon cooling. 
  	\textbf{(b)} Extracted peak positions of the four intrinsic CEF excitations as a function of temperature. 
  	The intrinsic transitions remain essentially temperature independent.
  	\textbf{(c)} Schematic CEF level scheme of Nd$^{3+}$ ($J=9/2$) in NdOF. 
  	The ground multiplet splits into five Kramers doublets under $C_{3v}$ symmetry, yielding four Raman-active transitions from the ground-state doublet to excited states at 1.7, 15.6, 19.2, and 80.9~meV.
  	}
  	\label{fig:figure2}
  \end{figure*} 
  
\textit{CEF excitations and nonlinear field splitting} --- {\color{black}In addition to phonons, Raman spectra reveal a series of extra peaks that we assign to CEF excitations of \ce{Nd^{3+}} [Fig.~\ref{fig:figure2}\textbf{(a)}]. 
Their assignment as CEF modes is supported by several considerations. 
First, powder XRD confirms the $R\overline{3}m$ structure remains stable down to 1.8~K [Fig.~\ref{fig:figure1}\textbf{(b)}], ruling out structural transitions as the origin of new phonons. 
Second, no magnetic ordering is detected down to 1.8~K, excluding collective magnon excitations. 
Third, the single-ion electronic configuration must be considered: for \ce{Nd^{3+}} the Hund's-rule ground-state multiplet is $^{4}I_{9/2}$, while the first excited multiplet $^{4}I_{11/2}$ lies about 250~meV ($\sim$2000~cm$^{-1}$) higher in energy~\cite{PhysRevB.45.10075,liu2006spectroscopic}. 
Thus all excitations observed below 200~meV must originate within the $^{4}I_{9/2}$ manifold. 
In $C_{3v}$  point-group symmetry, the $J=9/2$ multiplet is expected to split by the CEF interaction into five Kramers doublets, typically spread over an energy range of order 100~meV \cite{PhysRevB.88.104421,PhysRevB.89.134410,PhysRevB.91.224430,PhysRevResearch.6.043061,PhysRevLett.118.107202}. 
Consequently, four Raman-active CEF excitations from the ground-state doublet to higher-lying doublets are expected, exactly matching the four intrinsic peaks ($P_{\mathrm{CEF1}}$--$P_{\mathrm{CEF4}}$) observed here~{\color{black}[Fig.~\ref{fig:figure2}\textbf{(a)}]}. 
These intrinsic modes persist down to the lowest measured temperature of 1.7~K, whereas weaker extrinsic features arising from transitions between thermally populated excited states are significantly suppressed below 40~K and eventually vanish~[Fig.~\ref{fig:figure2}\textbf{(b)}]. 
The excitation energies of 1.7, 15.6, 19.2, and 80.9~meV (14, 126, 155, and 652~cm$^{-1}$) extracted from the spectra [Fig.~\ref{fig:figure2}\textbf{(c)}] are fully consistent with this CEF scheme and with earlier reports on related oxifluorides~\cite{holsa1998analysis}.}

{\color{black}To quantitatively account for the Raman-active CEF excitations in NdOF, we employ a CEF Hamiltonian constrained by the local $C_{3v}$ symmetry of the \ce{Nd^{3+}} site~\cite{hutchings1964point}:
\begin{equation}
	\hat{H}_{\text{CEF}} =  
	B_{2}^{0} \hat{O}_{2}^{0} + B_{4}^{0} \hat{O}_{4}^{0} + B_{4}^{3} \hat{O}_{4}^{3} 
	+ B_{6}^{0} \hat{O}_{6}^{0} + B_{6}^{3} \hat{O}_{6}^{3} + B_{6}^{6} \hat{O}_{6}^{6},
	\label{CEFHamiltonian}
\end{equation}
where $B_m^n$ are the CEF parameters and $\hat{O}_{m}^{n}$ the Stevens operators. 
The four CEF excitations observed at zero field establish the basic level scheme but are insufficient to uniquely fix all six independent $B_m^n$ parameters. 
To overcome this limitation, we analyze the field dependence of the CEF states. 
An applied magnetic field induces Zeeman splitting of the doublets, lifting degeneracies and shifting the excitation energies, thereby providing additional constraints that allow a unique determination of the CEF Hamiltonian.
}

\begin{figure}[h]
 	\centering
 	\includegraphics[width=\linewidth]{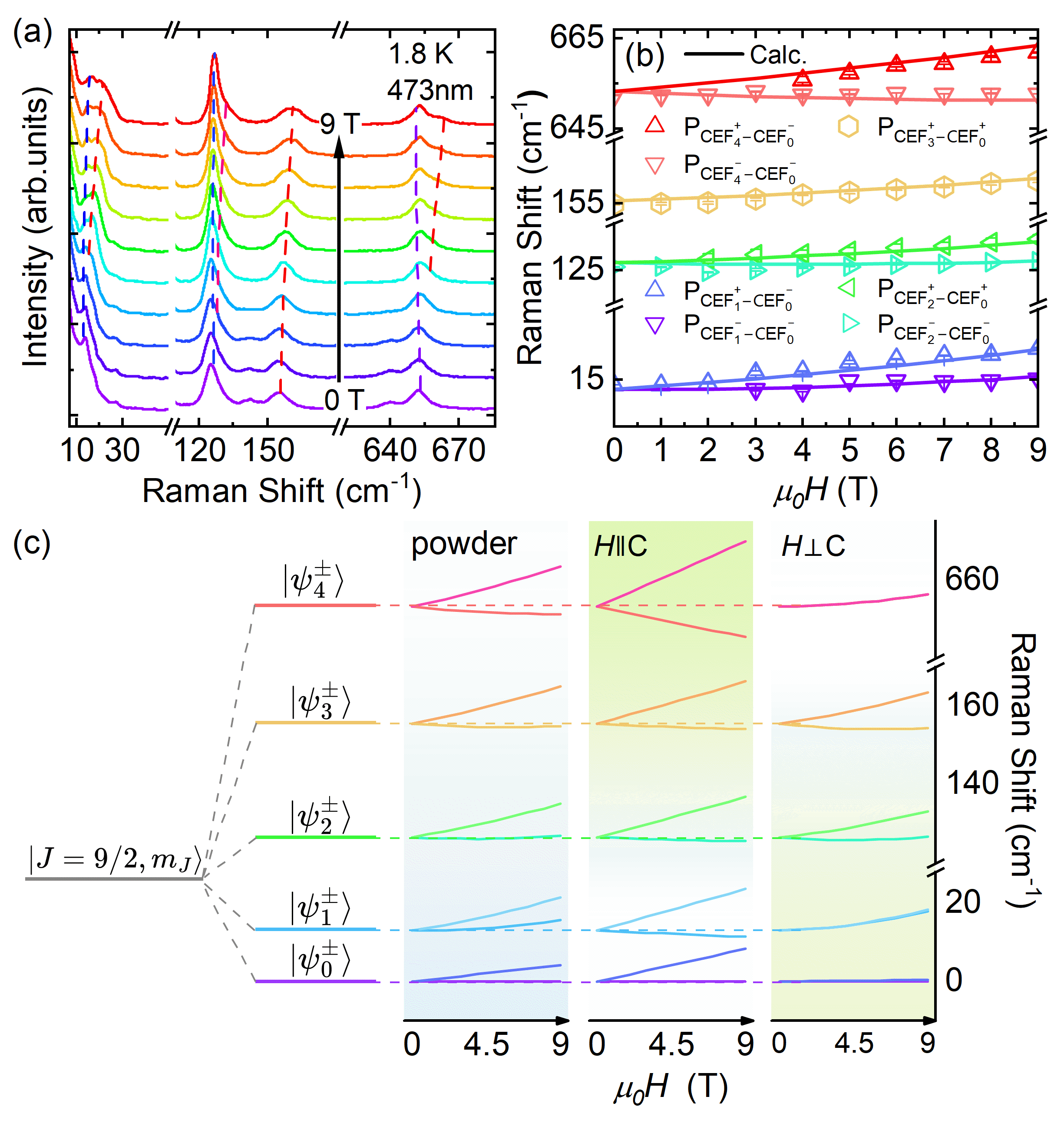}
 	\caption{ 
 	\textbf{Magnetic-field evolution of CEF excitations in NdOF. }
 	\textbf{(a)} Raman spectra measured at 1.8~K with 473~nm excitation under fields up to 9~T. 
 	The four intrinsic CEF peaks split and shift with increasing field, revealing multiple branches at high fields.  
 	\textbf{(b)} Extracted peak positions as a function of field. 
 	Seven distinct transitions are resolved, corresponding to excitations from the ground-state doublet to field-split excited states, in quantitative agreement with the CEF+Zeeman analysis.
 	\textbf{(c)}  Calculated Zeeman splitting of the $J=9/2$ manifold for powder-averaged samples, compared with single-crystal calculations for $H \parallel c$ and $H \perp c$. 
 	The results show nearly linear splitting for fields along $c$, whereas strongly nonlinear splitting arises for in-plane fields, consistent with the experimental powder response.
 	}
 	\label{fig:figure3}
 \end{figure}

{\color{black}As shown in Fig.~\ref{fig:figure3}(a), we performed Raman scattering measurements on NdOF powder under applied magnetic fields at 1.8~K. 
The four intrinsic CEF excitations exhibit clear field-induced splittings and shifts. 
To analyze these effects in a powder sample, we developed a dedicated scheme for modeling field-dependent CEF spectra. 
Specifically, we first calculate the CEF level splitting for a set of randomly chosen crystallographic orientations with respect to the external field. 
We then average the results over a sufficiently large number of orientations to obtain the powder-averaged response. 
By iteratively comparing the calculated spectra with the experimental data, we achieve a consistent determination of the CEF parameters.
The success of our powder-averaged CEF approach is highlighted in Fig.~\ref{fig:figure3}\textbf{(b)}. 
The symbols represent the experimentally extracted Raman peak positions, while the solid lines are calculated from the CEF Hamiltonian (Eq.~\ref{CEFHamiltonian}) including Zeeman splitting with powder averaging. 
The agreement is remarkable: the field evolution of the zero--field excitations is fully captured, with the original four CEF modes splitting into seven distinct branches under applied field. 
These branches correspond to the peaks $P_\mathrm{CEF_1^--CEF_0^-}$, $P_\mathrm{CEF_1^+-CEF_0^-}$,$P_\mathrm{CEF_2^--CEF_0^-}$,$P_\mathrm{CEF_2^+-CEF_0^+}$, 
$P_\mathrm{CEF_3^+-CEF_0^+}$, $P_\mathrm{CEF_4^--CEF_0^-}$, and $P_\mathrm{CEF_4^+-CEF_0^-}$. 
The quantitative consistency between experiment and theory confirms that the refined CEF parameters, constrained simultaneously by zero-field spectra and field-dependent splittings, provide a robust description of the \ce{Nd^{3+}} CEF scheme in NdOF.
The refined parameters, together with the corresponding wave functions and level scheme, are summarized in the SM~\cite{SuppInfo}.
}

{\color{black}An intriguing aspect of NdOF is that its CEF excitations exhibit pronounced nonlinear field dependence. 
To clarify the origin of this behavior, we systematically calculated the Zeeman splitting of the CEF doublets for different field orientations. 
As shown in Fig.~\ref{fig:figure3}(c), fields applied along the $c$-axis produce nearly linear shifts, whereas in-plane fields ($H \perp c$) lead to strongly nonlinear splittings due to enhanced mixing among closely spaced doublets. 
The nonlinear evolution observed in the powder-averaged spectra [Fig.~\ref{fig:figure3}(b)] therefore primarily reflects the response to in-plane magnetic fields.
}

\textit{Field-induced multipolar ground state} --- {\color{black}Beyond establishing the CEF level scheme, further insight can be gained 
from the symmetry analysis of the CEF wave functions. 
The ground-state and excited doublets of the $^{4}I_{9/2}$ multiplet are listed in the SM~\cite{SuppInfo}. 
In a $C_{3v}$ CEF, the five Kramers doublets split into irreducible representations $3\Gamma_{4}+2\Gamma_{5,6}$, consistent with group-theoretical expectations. 
Unlike Nd$_2$Zr$_2$O$_7$, which hosts a dipolar-octupolar ground state~\cite{PhysRevB.92.224430,PhysRevLett.115.197202,PhysRevB.91.174416,PhysRevB.92.144423,PhysRevB.92.184418}, NdOF realizes a purely dipolar doublet, separated from the first excited state by a small gap of 1.7~meV. 
In trigonal environments, axial distortions can substantially tune the CEF parameters, especially $B_{2}^{0}$, thereby driving a crossover between dipolar ($\Gamma_{4}$) and dipolar-octupolar ($\Gamma_{5,6}$) ground states~\cite{PhysRevLett.112.167203}. 
Point-charge model calculations of the Nd--O/F coordination polyhedron confirm that compressing or elongating the $C_{3}$ axis continuously modifies $B_{2}^{0}$, leading to such a transition at a critical distortion [see SM~\cite{SuppInfo}]. 
Thus, applying external pressure offers a promising route to tune the symmetry of the CEF ground-state wave function, providing a powerful means to control the magnetic properties of NdOF.
}

\begin{figure*}[t]
	\centering
	\includegraphics[width=\linewidth]{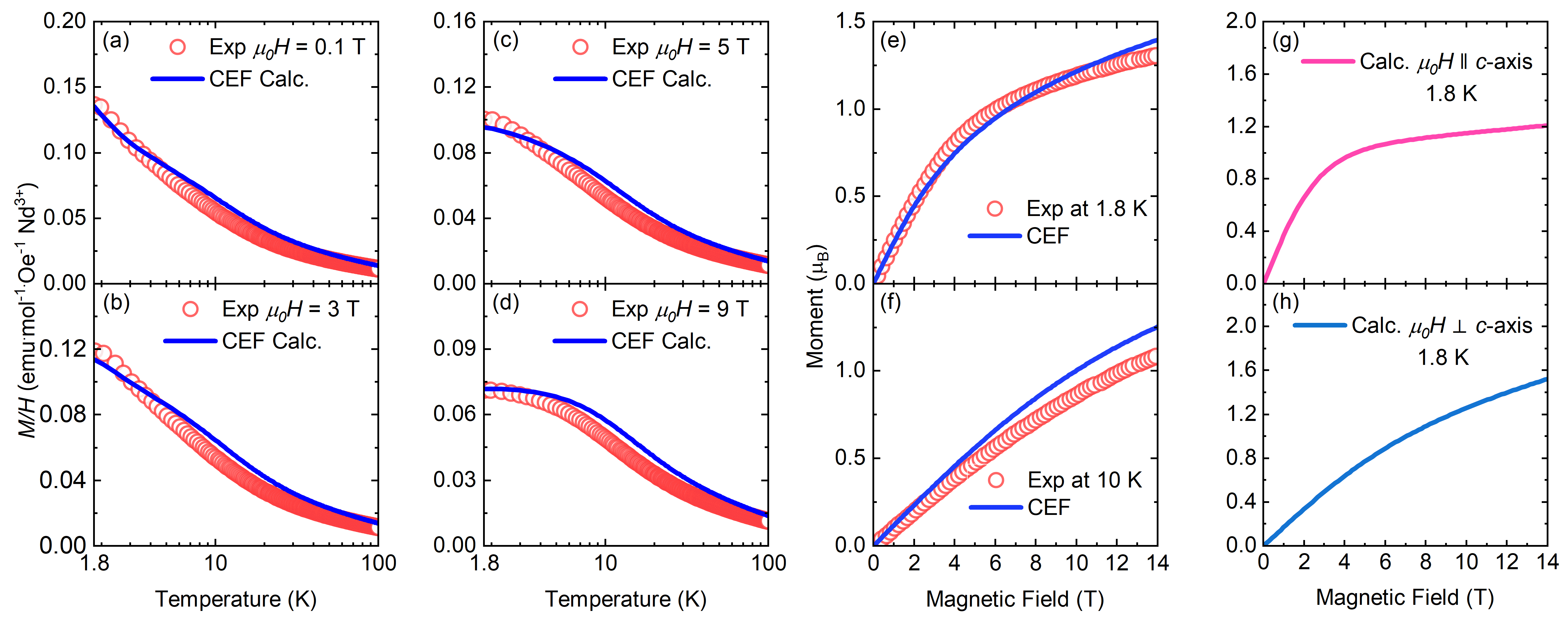}
	\caption{ \color{black}
	\textbf{$M/H$-$T$ and $M$-$H$ of NdOF and comparison with CEF modeling.}
	\textbf{(a) -- (d)} $M/H$-$T$ under fields of 0.1, 3, 5, and 9~T.  
	The CEF simulations (blue solid lines) reproduce the experimental data (symbols) over the entire range.  
	\textbf{(e), (f)} $M$-$H$ at 1.8 and 10~K. 
	The CEF model (blue solid lines) is in good agreement with experiment.  
	\textbf{(g), (h)}  Calculated CEF magnetization for fields along the $c$-axis and within the $ab$-plane at 1.8~K.
	}
	\label{fig:figure4}
\end{figure*}  

{\color{black}Beyond structural tuning, external magnetic fields hold promise as an additional effective means to manipulate the CEF ground state. 
As discussed above, axial compression or elongation of the Nd--O/F polyhedron can drive a transition between purely dipolar and dipolar-octupolar doublets by modifying the $B_{2}^{0}$ parameter. 
Remarkably, our data suggests a similar effect may potentially be achieved by applying a magnetic field.}
{\color{black}As shown in Fig.~\ref{fig:figure4}\textbf{(a)-(d)}, the  $M/H$-$T$ exhibits a pronounced field dependence: while a Curie--Weiss-like upturn is present at low fields, a broad plateau develops with increasing external field strength in the low-temperature regime.
This crossover is further reflected in the $M$-$H$ measurements.  
As displayed in Fig.~\ref{fig:figure4}\textbf{(e)} and \textbf{(f)}, the magnetization at 1.8~K grows rapidly at low fields and approaches saturation 
above $\sim 7$~T, in contrast to the more gradual evolution at elevated temperatures.}
{\color{black}Considering that high magnetic fields break the time-reversal symmetry of Kramers ions and fundamentally modify the symmetry of their CEF wave functions, the ground state doublet, initially of purely dipolar character, inevitably admixes excited-state components, thereby giving rise to a ferromagnetic state with pronounced multipolar character.}
{\color{black}To further validate the CEF scheme, we calculated the powder-averaged magnetization using the CEF parameters extracted from the Raman data. 
The powder average takes the standard form for a randomly oriented polycrystalline sample: $\chi_{\text{avg}} = \frac{1}{3} \chi_{\parallel c} + \frac{2}{3} \chi_{\perp c}$. 
As shown by the solid lines in Fig.~\ref{fig:figure4}\textbf{(a)}-\textbf{(d)} curves are in reasonable agreement with the experimental data across the field range of 0.1--9~T, confirming the reliability of the extracted CEF parameters. 
The corresponding $M$-$H$ curves are also well reproduced, as shown in Fig.~\ref{fig:figure4}\textbf{(e)} and \ref{fig:figure4}\textbf{(f)}.
Nevertheless, some discrepancies remain, particularly in the low-temperature and low-field regime, where the influence of inter-ion exchange interactions is expected to become relevant. 
In general, the complete Hamiltonian for NdOF incorporating the CEF, anisotropic exchange interactions, and the Zeeman effect can be written as:}
\begin{equation}
	\begin{aligned}
		\hat{H}  =& \hat{H}_{\text{CEF}} + \sum_{\langle i j\rangle} \left[ \vartheta_{zz} J_{i}^{z} J_{j}^{z} + \vartheta_{\pm} (J_{i}^{+} J_{j}^{-} + J_{i}^{-} J_{j}^{+}) \right. \\
		& \left. + \vartheta_{\pm \pm} (\gamma_{ij} J_{i}^{+} J_{j}^{+} + \gamma_{ij}^{*} J_{i}^{-} J_{j}^{-}) \right. \\
		& \left.+ \vartheta_{z \pm} (\gamma_{ij} J_{i}^{+} J_{j}^{z} + \gamma_{ij}^{*} J_{i}^{-} J_{j}^{z} + \langle i \leftrightarrow j\rangle) \right] \\
		&- \mu_0 \mu_B g_J \sum_i (h_x \hat{J}_i^x + h_y \hat{J}_i^y + h_z \hat{J}_i^z)
	\end{aligned}
	\label{eq:Hamiltonian}
\end{equation}
{\color{black}where $\vartheta_{zz}$ and $\vartheta_{\pm}$ denote the nearest-neighbor anisotropic exchange parameters along the longitudinal and transverse channels, respectively. 
The off-diagonal terms $\vartheta_{z\pm}$ and $\vartheta_{\pm\pm}$ capture the bond-dependent anisotropy inherent to the honeycomb lattice. 
Within a mean-field treatment, these bond-dependent components cancel upon summing over the three nearest-neighbor bond directions (SM). 
The last term represents the Zeeman coupling of the total angular momentum operators $\hat{J}$ to the external field.}

{\color{black}However, reliably extracting the exchange parameters from powder magnetization data alone is challenging for NdOF, given the small magnitude of the exchange couplings relative to the CEF energy scales. Accurately determining these parameters will require the growth of single crystals and the combination of anisotropic magnetization measurements with momentum-resolved spectroscopic probes such as inelastic neutron scattering. 
We leave this as an important direction for future work.}

{\color{black}While a precise determination of the exchange parameters awaits future single-crystal studies, the CEF physics alone already reveals a striking feature of NdOF at high magnetic fields: the ground-state wave function undergoes a continuous reconstruction that can be directly tracked by analyzing its symmetry composition.}
{\color{black}We systematically calculated the ground state wave functions at 3~T, 5~T, 7~T, and 9~T field applied along the $a$-axis where exchange interactions are negligible and the CEF contribution dominates the magnetization, quantifying the weights of their dipolar-octupolar components (Table~\ref{tab:tableI}). These results clearly demonstrate that the admixture from the first excited state ($\Gamma_{5,6}$) continuously increases with magnetic field strength, systematically tuning the degree of dipolar-multipolar mixing in the ground state. This continuously tunable evolution substantiates the potential of external magnetic fields as an effective means to continuously manipulate the multipolar character in rare-earth magnets.}

\begin{table*}
	\centering
	\caption{Weights of dipolar-octupolar moments under different magnetic fields}
	\renewcommand{\arraystretch}{1.5}
	\centering
	\begin{tabular}{c p{11cm} c} 
		\hline
		\hline
		Magnetic Field (T) & \multicolumn{1}{c}{Wave Function}  &  Ratio \\
		\hline
		
		3 & 
		$\begin{array}{l}
			\left| \psi_{0}^{-} \right\rangle = 
			\quad 0.0221 \left| -\tfrac{9}{2} \right\rangle 
			+ 0.4403 \left| -\tfrac{7}{2} \right\rangle 
			+ 0.3003 \left| -\tfrac{5}{2} \right\rangle 
			- 0.0349 \left| -\tfrac{3}{2} \right\rangle \\
			\quad \quad \quad\quad
			- 0.4025 \left| -\tfrac{1}{2} \right\rangle 
			+ 0.4590 \left| +\tfrac{1}{2} \right\rangle  
			+ 0.1324 \left| +\tfrac{3}{2} \right\rangle 
			+ 0.2704 \left| +\tfrac{5}{2} \right\rangle \\ 
			\quad \quad \quad\quad
			+ 0.4977 \left| +\tfrac{7}{2} \right\rangle 
			+ 0.0574 \left| +\tfrac{9}{2} \right\rangle
		\end{array}$
		& 2\% \\
		
		5 &  
		$\begin{array}{l}
			\left| \psi_{0}^{-} \right\rangle = 
			\quad - 0.0239 \left| -\tfrac{9}{2} \right\rangle 
			- 0.3788 \left| -\tfrac{7}{2} \right\rangle 
			- 0.3223 \left| -\tfrac{5}{2} \right\rangle 
			+ 0.0298 \left| -\tfrac{3}{2} \right\rangle \\
			\quad \quad \quad\quad
			+ 0.3357 \left| -\tfrac{1}{2} \right\rangle 
			- 0.4907 \left| +\tfrac{1}{2} \right\rangle  
			- 0.2140 \left| +\tfrac{3}{2} \right\rangle 
			- 0.2423 \left| +\tfrac{5}{2} \right\rangle \\ 
			\quad \quad \quad\quad
			- 0.5331 \left| +\tfrac{7}{2} \right\rangle 
			- 0.0941 \left| +\tfrac{9}{2} \right\rangle
		\end{array}$
		& 6\% \\
		
		7 & 
		$\begin{array}{l}
			\left| \psi_{0}^{-} \right\rangle = 
			\quad - 0.0120 \left| -\tfrac{9}{2} \right\rangle 
			- 0.2925 \left| -\tfrac{7}{2} \right\rangle 
			- 0.3432 \left| -\tfrac{5}{2} \right\rangle 
			- 0.0048 \left| -\tfrac{3}{2} \right\rangle \\
			\quad \quad \quad\quad
			+ 0.2437 \left| -\tfrac{1}{2} \right\rangle 
			- 0.5263 \left| +\tfrac{1}{2} \right\rangle  
			- 0.2837 \left| +\tfrac{3}{2} \right\rangle 
			- 0.2016 \left| +\tfrac{5}{2} \right\rangle \\ 
			\quad \quad \quad\quad
			- 0.5682 \left| +\tfrac{7}{2} \right\rangle 
			- 0.1268 \left| +\tfrac{9}{2} \right\rangle
		\end{array}$
		& 10\% \\
		
		9 & 
		$\begin{array}{l}
			\left| \psi_{0}^{-} \right\rangle = 
			\quad 0.0082 \left| -\tfrac{9}{2} \right\rangle 
			- 0.2042 \left| -\tfrac{7}{2} \right\rangle 
			- 0.3542 \left| -\tfrac{5}{2} \right\rangle 
			- 0.0561 \left| -\tfrac{3}{2} \right\rangle \\
			\quad \quad \quad\quad
			+ 0.1486 \left| -\tfrac{1}{2} \right\rangle 
			- 0.5495 \left| +\tfrac{1}{2} \right\rangle  
			- 0.3354 \left| +\tfrac{3}{2} \right\rangle 
			- 0.1606 \left| +\tfrac{5}{2} \right\rangle \\ 
			\quad \quad \quad\quad
			- 0.5866 \left| +\tfrac{7}{2} \right\rangle 
			- 0.1524 \left| +\tfrac{9}{2} \right\rangle
		\end{array}$
		& 14\% \\
		
		\hline
		\hline
	\end{tabular}
	
	\label{tab:tableI}
	
\end{table*}
The field thus acts as a continuous control knob, transforming the ground state from a purely dipolar doublet to one with mixed dipolar-multipolar nature. Physically, the application of a magnetic field --- particularly within the $ab$-plane --- fundamentally alters the symmetry of the ground-state CEF wave function, introducing substantial multipolar components.
This field-induced reconstruction of the ground state explains the emergence of the nearly temperature-independent susceptibility plateau at high fields.
The corresponding calculations for single-crystal orientations [Fig.~\ref{fig:figure4}\textbf{(g)} and \ref{fig:figure4}\textbf{(h)}] further reveal the anisotropic response: while fields along the $c$-axis drive a nearly linear approach to saturation, in-plane fields produce a more gradual evolution.
At sufficiently high fields the system enters a polarized state. 
Importantly, this polarization is not a simple dipolar ferromagnet, but rather a field-induced state in which the ground-state wave functions acquire multipolar components, leading to a mixed dipolar-multipolar form of ferromagnetic polarization.
{\color{black}Our results highlight a general mechanism by which the ground-state doublet evolves to acquire significant multipolar character under external fields, a principle that may apply broadly to rare-earth honeycomb magnets.}

\textit{Summary} ---
{\color{black}NdOF, a rare-earth chalcohalide with SmSI-type structure and local $C_{3v}$ symmetry at the Nd site, provides a clean platform to study field-tunable CEF physics. 
Polarized Raman spectra resolve four intrinsic CEF excitations at 1.7, 15.6, 19.2, and 80.9~meV, while additional thermally activated processes are significantly suppressed below 40~K and eventually vanish. 
Using a powder-averaged Zeeman--CEF approach, we quantitatively reproduce the nonlinear field-induced splitting of these modes into seven distinct branches, most pronounced for in-plane fields. 
$M/H$-$T$ and $M$-$H$ measurements reveal a crossover from Curie--Weiss behavior at low fields to a nearly temperature-independent plateau at high fields, {\color{black}which is well described by the CEF model for the $J = 9/2$ manifold. }
Together these results demonstrate that the NdOF ground state, initially a predominantly dipolar doublet, can be continuously evolved under external fields or lattice distortions to acquire multipolar components.}

\begin{acknowledgments}
This work was supported by the National Key Research and Development Program of China (Grant Nos. 2022YFA1402704 and 2024YFA1408300), the National Science Foundation of China (Grant No. 12274186), the Strategic Priority Research Program of the Chinese Academy of Sciences (Grant No. XDB33010100), the CAS Superconducting Research Project under Grant No. SCZX-0101, the Fund for High-level Talents of Lanzhou and the Synergetic Extreme Condition User Facility (SECUF) \href{https://cstr.cn/31123.02.SECuf}{SECuf}.
\end{acknowledgments}
		
%

\end{document}